\documentclass[twocolumn,amsfonts,amsmath,amssymb,final,longbibliography,aps,prd,footinbib,nofootinbib,10pt]{revtex4-1}
\usepackage[dvipsnames]{xcolor}
\usepackage{dcolumn}
\usepackage{lipsum}
\usepackage[colorlinks=true,allcolors=Blue]{hyperref}
\usepackage{upgreek,graphicx,manfnt,pifont,colonequals,bbm,stmaryrd}
\usepackage[shortlabels]{enumitem}
\usepackage{mathtools}

\usepackage{amstext}
\usepackage{dsfont} 

\def\be{\begin{eqnarray}}
\def\ee{\end{eqnarray}}
\def\bea{\begin{eqnarray}}
\def\eea{\end{eqnarray}}

\newcommand{\Id}{\mathds{1}}
\newcommand{\cg}{\textsl{g}}

\newcommand{\beal}{\begin{equation}
\begin{aligned}}
\newcommand{\eeal}{\end{aligned}
\end{equation}}
\newcommand{\bem}{\begin{multline}}
\newcommand{\eem}{\end{multline}}

\begin{document}

\title{On a categorical structure of the set of all CFTs}

\medskip

\author{Rotem Ben Zeev, Behzat Ergun, Elisa Milan, and Shlomo S. Razamat}
 \email{rotemitbz@gmail.com, behzat.ergun@campus.technion.ac.il, elisa.milan90@gmail.com, razamat@physics.technion.ac.il}

 \medskip

\affiliation{Department of Physics, Technion, Haifa 32000, Israel}

 \date{\today}


\begin{abstract}
\noindent  We identify a categorical structure of the set of all CFTs. In particular, we show that the set of all CFTs has a natural monoidal strict $2$-category structure with the $1$-morphisms being sequences of deformations and $2$-morphisms determined by $0$-form symmetries of the CFTs.  
\end{abstract}

\maketitle

\section{Introduction}

Conformal field theories (CFTs) are basic mathematical structures in the study of  a wide range of physical phenomena. A plethora of different CFTs 
arises when one studies physical systems in various dimensions and in various contexts. CFTs are also  a fruitful ground to study general mathematical  properties of quantum physics. An interesting question  is to understand whether one can classify all possible CFTs and their properties: in other words,  what is the ``space'' of all CFTs and does it have a natural mathematical structure?   

One can  approach this question in different ways. One such approach, based on assuming some properties of a theory of interest and fully exploiting constraints following from conformal symmetry to deduce additional properties, is called the {\it conformal bootstrap}. This approach focuses on  a single given CFT and has received a lot of attention in recent years \cite{Hartman:2022zik,Poland:2022qrs}. One of the basic properties defining a CFT (and a quantum field theory (QFT) in more generality) is the notion of symmetry. We can think of symmetry as a collection of topological extended operators a given theory admits \cite{Gaiotto:2014kfa}. This turns out to be a very rich structure associating mathematically a higher fusion category to a given theory. For an elementary introduction to categories see {\it e.g.} \cite{MR1712872}. The higher fusion category generalizes the familiar notion of a group one associates to a $0$-form symmetry. The generalized structure  incorporates anomalies, higher form symmetries, higher group structures, and non-invertible topological defects into a single rich structure. Understanding and fully exploiting such structures also has received a lot of attention recently (see {\it e.g.} \cite{Cordova:2022ruw,Freed:2022qnc}). Studying symmetries of a theory, in principle, also focuses on properties of a given theory of interest. However, the symmetry structure of a theory obtained as a deformation of another one is constrained by the latter through the ideas of matching anomalies following the seminal work of `t Hooft \cite{tHooft:1979rat}. (See {\it e.g.} \cite{Komargodski:2020mxz, Cordova:2019jqi} for more modern applications of these ideas.)

In this note, we want to consider a connection between deformations and higher fusion categories of topological defects. In particular, we will discuss how one can understand more mathematically the set of all CFTs, either in a given dimension or in any dimension, as a category. More precisely we will discuss a $2$-category structure of the set of CFTs.\footnote{Categorical structure of the space of CFTs was discussed before, see {\it e.g.} \cite{Gaiotto:2015aoa}. There, the morphisms are taken to be interfaces of two CFTs. One might try and relate  the picture of RG domain walls of \cite{Gaiotto:2012np,Dimofte:2013lba} to the one we present here (We thank C.~Beem and Y.~Tachikawa for stressing this to us.).
Moreover, TQFTs (and CFTs) themselves can be defined as functors between various categories \cite{MR2079383,MR1001453,Kontsevich:2021dmb,Dedushenko:2022zwd}. Categorical language to organize various conjectures about class ${\cal S}$ \cite{Gaiotto:2009hg,Gaiotto:2009we} theories were discussed in \cite{Tachikawa:2017byo}. } For an extensive introduction to 2-categories see {\it e.g.}~\cite{Gray1974-qy, Johnson2020-oa}.\footnote{A rigorous definition of 2-categories with monoidal structure can be found in \cite{Kapranov-book, KAPRANOV1994241}. See also \cite{Ahmadi2020-lq, BAEZ1996196}.} The $2$-category will have objects given by the CFTs, $1$-morphisms related to deformations taking one CFT to another, and finally $2$-morphisms related to $0$-form  symmetry. Once this is done, one can, in principle, add to the discussion  the higher form symmetries, and more generally higher categorical structures,  though we will refrain from doing that explicitly here. The category structure of the set of CFTs would, in principle, impose certain mathematical relations on the space of QFTs. {\it The main motivation for this note is to rewrite known facts about CFTs in the categorical language with the hope that this will lead eventually to deeper insights into the structure of the set of all CFTs.}

Before we begin let us stress the general philosophy of the construction. We will start with the set of CFTs and will study relations between these given by various deformations. Importantly, we will not study {\it all} deformations of a given CFT. A general deformation can lead to a variety of behaviors at low energy (IR): for example, one can obtain free vector fields in the IR. The study of general structure of RG flows is an interesting problem, see {\it e.g.} \cite{Gukov:2016tnp}: we will only consider deformations which take a CFT to a CFT.  

The outline of the note is as follows. We will first discuss the $1$- and $2$-category structure of set of CFTs in given dimension $D$. Then we will discuss the categorical structure of the set of CFTs in any dimension. Finally, we will make several comments and translate some of the physical statements and conjectures about the space of all possible CFTs to this categorical language.

\

\section{Category of CFTs}
Let us define the set ${\cal C}^{(D)}_0$ to be the set of all CFTs  in $D$ space-time dimensions. By a CFT we mean a unitary quantum field theory which has conformal symmetry.\footnote{We will not distinguish here between relative and absolute CFTs 
\cite{Freed:2012bs}. In principle we can also consider the CFTs to be defined with the corresponding symmetry TFT \cite{Freed:2012bs,Gaiotto:2020iye,Apruzzi:2021nmk,Freed:2022qnc}. Moreover, one in principle could also consider {\it direct sums} of theories \cite{Sharpe:2022ene} ({\it e.g.} these naturally seem to arise in some compactification scenarios \cite{Gadde:2015wta}): we will not explicitly consider these here.} The action of the conformal symmetry on various structures in the theory can be nontrivial (faithful) if the theory is a proper CFT, or can be non-faithful in the degenerate, TQFT, case.
One can further specialize in various ways. For example, we can consider only CFTs with particular amount of supersymmetry, or CFTs residing on the same conformal manifold. We will do so later on but for now, we shall keep the discussion more generic.

\begin{figure}
\includegraphics[scale=0.25]{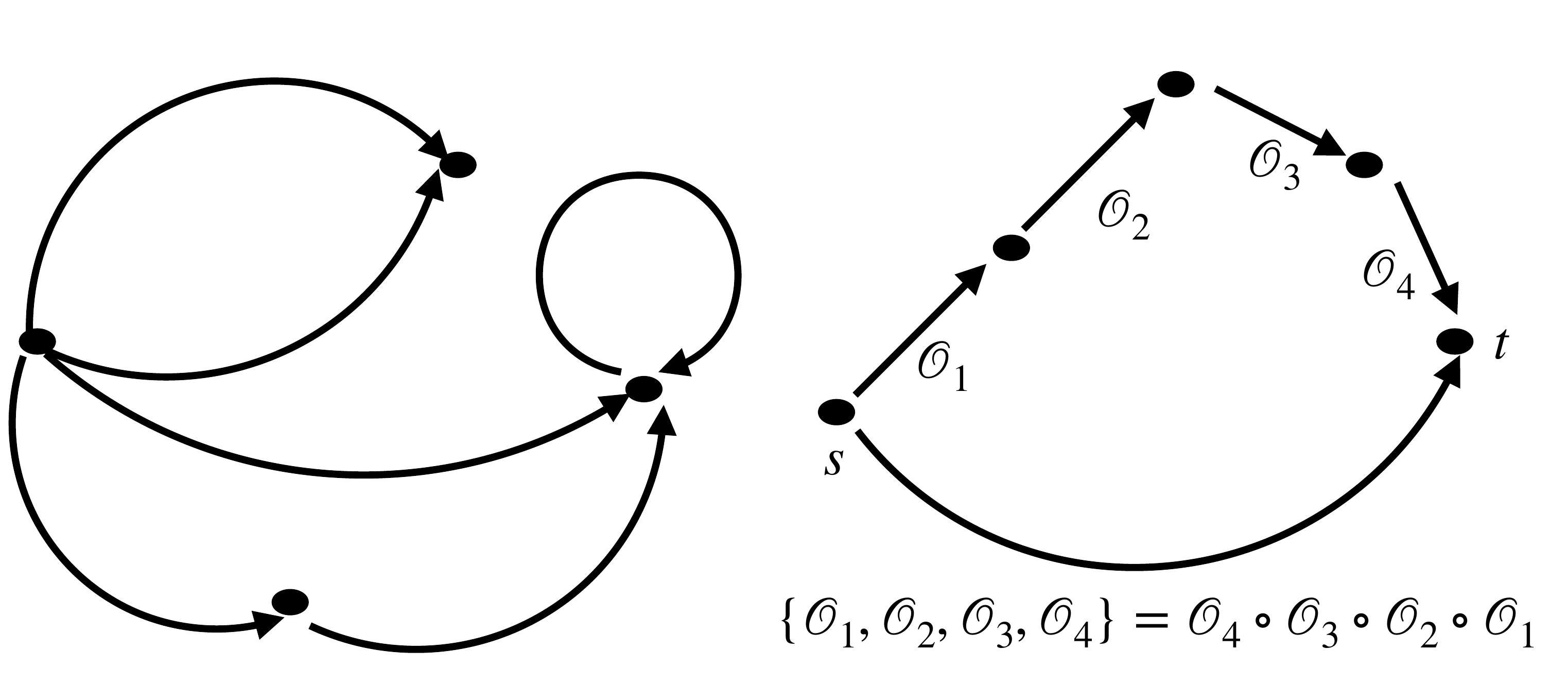}
\caption{ \label{cat}The category of CFTs and the morphisms as sequences of deformations.}
\end{figure}

We define a $1$-category ${\cal C}^{(D)}$ so that its objects are given by ${\cal C}^{(D)}_0$. A morphism connecting two objects $X,\, Y\in {\cal C}^{(D)}_0$, $f:X\to Y$, corresponds to a field theoretic operation on CFT $X$ which results in CFT $Y$ in the {\it IR}. We will refer to the UV theory as the {\it source} and the IR CFT as the {\it target} of the deformation. An example of such  is performing a deformation corresponding to operator deformation ${\cal O}$. That is the correlation functions of the source and the target CFT are related schematically as,
\be 
\langle \,\dots\, \rangle_Y = \langle \,\dots\, e^{i\lambda\, \int d^Dx \,\cal O}\rangle_X\,,
\ee where on the right, we take the proper low energy limit of the correlators. If $X\neq Y$ the deformation will be either relevant or exactly marginal by definition.
If $X=Y$ we can have nontrivial morphisms which correspond to irrelevant deformations. The parameter $\lambda$ is the coupling. Note that in case of exactly marginal deformations different values of coupling are different morphisms with target objects being different.  In  the case of relevant and irrelevant deformation the precise magnitude of $\lambda$ is inessential (as long as it is of definite sign and is small enough) as this sets an RG scale while we are only interested in the CFT endpoints of the flow. Another type of morphism that we will consider is taking a CFT and gauging part of its (generalized) global symmetry.\footnote{The gauging, again, might depend on inessential continuous coupling constants or on consequential discrete parameters, such as the level of a Chern-Simons term in $3d$. We can also gauge discrete sub-groups of the global symmetry though we will focus the discussion on continuous ones.}
Finally, one, in principle, can also consider deforming a CFT by turning on vacuum expectation value (VEV) to an operator, but we will only discuss operator and gauging deformations in what follows.  Let us refer to these deformations as being the {\it basic} ones.  We will soon define the morphisms more rigorously.

We immediately deduce that the above definition of morphisms has to be extended for the structure to define a category. The issue is  completeness under composition of morphisms. Given objects $X,\, Y$, and $Z$ corresponding to CFTs, if we have two morphisms,
\be
f:X\to Y\,,\qquad g:Y\to Z\,,
\ee what is the morphism $g\circ f:X\to Z$ corresponding to the composition of the two? Naively we might be tempted to construct this morphism by searching for an appropriate deformation of one of the two types discussed above: deforming the CFT $X$, say, by an operator or gauging, leading to CFT $Z$. However, these are not enough to cover all the possibilities. Imagine that we go from $X$ to $Y$ using one of the deformations above ($f$) and the global symmetry of $Y$ is larger than that of $X$: some of the symmetry emerges in the IR. Then, as deformation $g$, we gauge a subgroup of this {\it emergent} symmetry. We cannot perform this operation directly on $X$ without first performing $f$, flowing to the IR and then gauging. Thus, in order for the structure we discuss to be a  well defined category we need  also to consider deformations of $X$ which are defined by any sequence of the basic deformations. See Figure \ref{cat}.

Let us be more precise and define the following. We consider, for concreteness, the set of morphisms between two  CFTs corresponding to operator deformations. Since the source theory is a CFT, we can classify all Lorentz scalar deformations by their scaling dimensions. Let us then 
consider the collection of all deformations of given scaling dimension $\Delta$, ${\cal O}_\alpha$.\footnote{One can generalize this discussion by turning on deformations of different scaling dimensions.} These deformations are in a (possibly reducible) representation of the $0$-form global symmetry group $G^{(X)}_0$ of the source CFT $X$. Namely, given a group element  $g\in G^{(X)}_0$ we have $g\cdot {\cal O}_\alpha={\cal O}_{g(\alpha)}$, meaning that acting on the deformation of given scaling dimension, we obtain another deformation of the same scaling dimension.  
Some morphisms, the basic ones, thus correspond to operators ${\cal O}_\alpha$. This is not the most general deformation: 
A general morphism is  defined by a sequence of basic deformations, an ordered tuple  $$\{{\cal O}_1,\,{\cal O}_2,\,\cdots , {\cal O}_n\}\equiv {\cal O}_n\circ \cdots \circ {\cal O}_2\circ {\cal O}_1\,,$$
such that we first flow with ${\cal O}_1$, then we deform the IR CFT with ${\cal O}_2$, and so on. 
Although we defined the above with operator deformation we can also consider gauging a subgroup of $G^{(X)}_0$ as one of the deformations.
The order of the deformations might matter. In particular, in some cases the sequence of deformations only makes sense in a particular order.
For example, we might want to gauge a symmetry which only emerges after we deform the source theory.

\section{Higher category structure}

\begin{figure}
\includegraphics[scale=0.2,trim={5px 200px 5px 200px},clip]{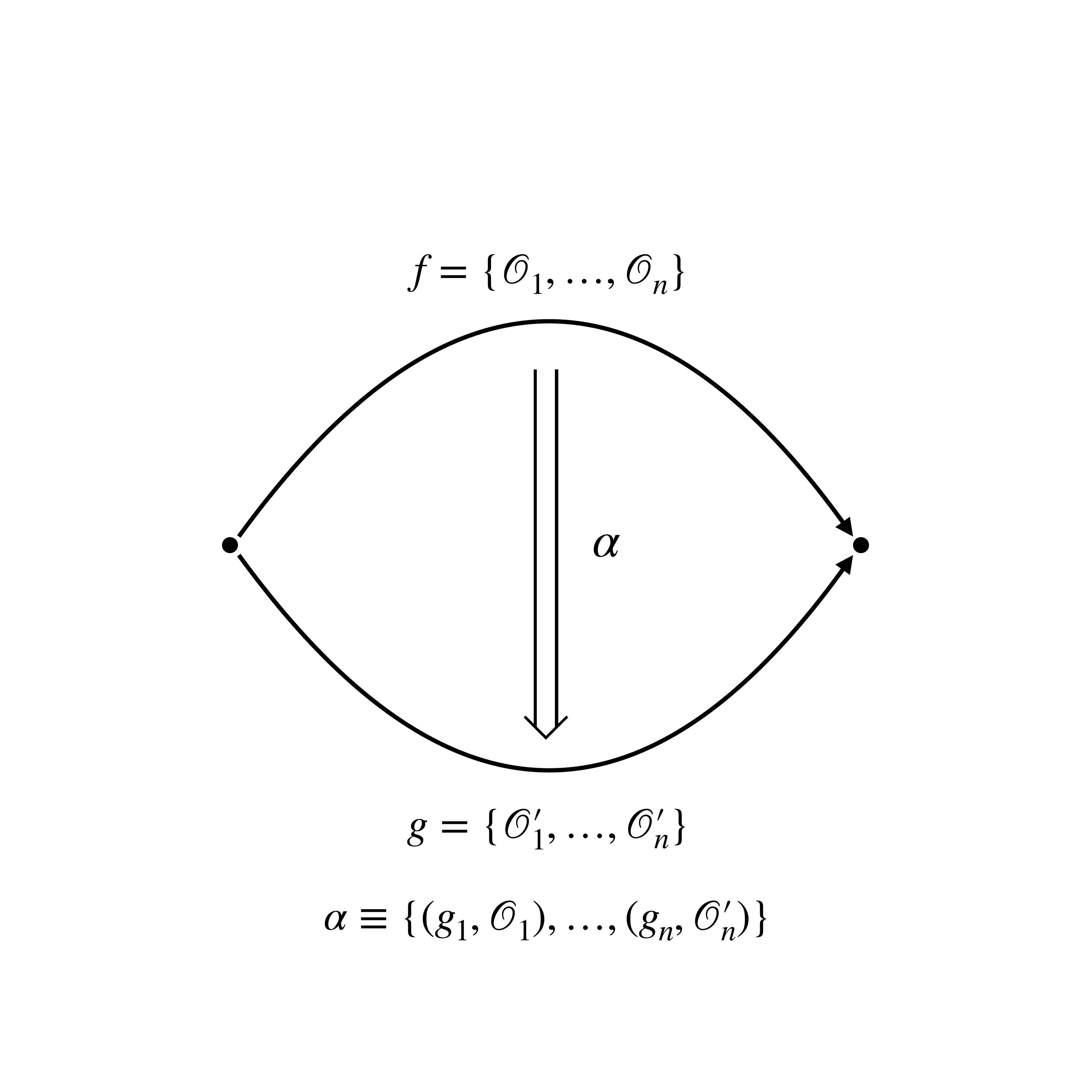}
\caption{ \label{2cat}The $2$-morphism as a sequence of group elements.}
\end{figure}

We can think of the group elements of $G^{(X)}_0$, or more precisely certain equivalence classes to be defined soon, as generating $2$-morphisms connecting different $1$-morphisms. 
The sequence of the CFTs appearing in the definition of a morphism has the following sequence of $0$-form symmetries,
$$\{G^{(X_1)}_0,\,G^{(X_2)}_0,\,\cdots\}\,.$$ We remind that the deformations can be operator deformations or gauging of symmetries. The symmetries in the sequence do not have to be subgroups (or quotients) of previous ones as some global symmetry can emerge in the IR. Moreover, even if part of the global symmetry of the source CFT is unbroken by the deformation, it can act trivially on target CFT. Given two different morphisms $f=\{{\cal O}_1, \ldots, {\cal O}_n\}:X_1 \rightarrow \cdots \rightarrow X_{n+1}$ and $f'=\{{\cal O}'_1, \ldots, {\cal O}'_n\}:X_1 \rightarrow \cdots \rightarrow X_{n+1}$ corresponding to the same sequence of CFTs, we might be able to define a $2$-morphism between them, denoted by $\alpha : f \Rightarrow f'$ 
in the following manner.
A $2$-morphism is an ordered set of 
pairs, each consisting of a group element and a source operator (see Figure \ref{2cat}),
\be
\label{eq: 2-morph}
\alpha \equiv \{ (g_1, {\cal O}_1),\,(g_2, {\cal O}_2),\,\dots,\,(g_n, {\cal O}_n)\}\,,
\ee
such that,
\be
g_i \in G_0^{(X_i)} \; \text{ and } \; {\cal O}'_i = g_i \cdot {\cal O}_i \; \text{ for each } i=1, \dots, n \,,
\ee
{\it i.e.}~in each pair the group element transforms a deformation in $f$ into the corresponding one in $f'$. This definition ensures that every $2$-morphism uniquely specifies the source and target 1-morphisms, as required by the axioms of $2$-categories. Note that we do not assume that any two $1$-morphims connecting the same objects are related by a $2$-morphism: the corresponding deformations might not be related by an action of the $0$-form symmetry.
As each deformation ${\cal O}_i$ might preserve some subgroup $H_i\subset G^{(X_i)}_0$,\footnote{The subroup $H_i= Stab_{G_0^{(X_i)}}({\cal O}_i)$ preserves a deformation $({\cal O}_i)_\alpha$ if for every $h\in H_i$ $\Rightarrow$ $({\cal O}_i)_{h(\alpha)}=({\cal O}_i)_\alpha$.}  
to be more precise we should replace each group element $g_i$ by the left coset $g_i H_i$, {\it i.e.} $\alpha = \{(g_1 H_1, {\cal O}_1)\,, \ldots, (g_n H_n, {\cal O}_n) \} $. Note that the identity $2$-morphism $Id_f$ on $f$ is just $Id_f= \{(H_1, {\cal O}_1),\,\ldots,\,(H_n, {\cal O}_n)\}$\footnote{Here, we do the following identification. If for some $i$, we have $X_i=X_{i+1}$ and ${\cal O}_i=id_{X_i}$ {\it i.e.} {\it no deformation}, then $H_i= G_0^{(X_i)}$, and we remove the $i$th entry from the sequence of $1$- and $2$-morphisms. This physically means we do nothing at step $i$.}.
In order not to clutter notations, we will keep labeling the equivalence classes by representative group elements.

\begin{figure}
\includegraphics[scale=0.2,trim={5px 200px 5px 200px},clip]{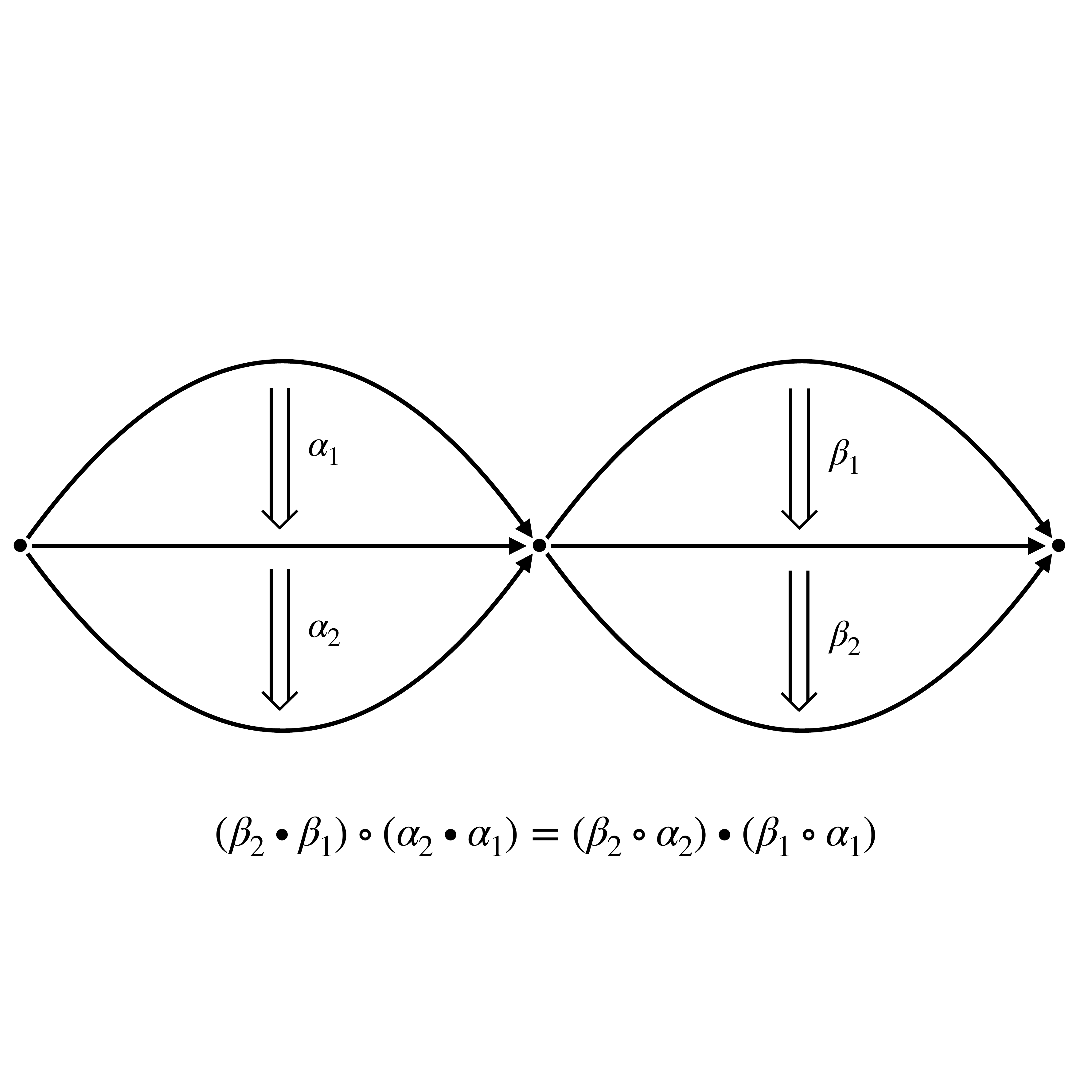}
\caption{ \label{inter}The interchange law.}
\end{figure}

 The definition of 2-category requires the presence of two different compositions (vertical and horizontal) for $2$-morphisms, satisfying a constraint called {\it interchange law}. In our construction, they are defined as follows. 
%
Given two $2$-morphisms $\alpha_1$ and $\alpha_2$
\beal
\alpha_1 &=\{(g_1H_1, {\cal O}_1),\ldots,\, (g_n H_n, {\cal O}_n)\,\}\,, \\ 
\alpha_2 &=\{(h_1 K_1, {\cal O}'_1),\ldots,\, (h_n K_n, {\cal O}'_n)\,\}\,,
\eeal
the vertical composition is given naturally using the group multiplication,\footnote{Note that, as a necessary condition for the vertical composition of $\alpha_1$ and $\alpha_2$ to be defined, we need that ${\cal O}'_i = g_i \cdot {\cal O}_i$ for some $g_i \in G_0^{(X_i)}$ for each $i$ in the sequence. In particular, these exist only if the sequence of CFTs in both $2$-morphisms is identical.}
\beal
\alpha_2 \bullet \alpha_1 & = \{(h_1 K_1 g_1 H_1, {\cal O}_1),\,\ldots,\, (h_n K_n g_n H_n, {\cal O}_n)\} \\
& =\{(h_1 g_1 H_1, {\cal O}_1),\,\cdots,\, (h_n g_n H_n, {\cal O}_n)\}\,
\eeal
where the last line follows since we have the relation $K_i = g_i H_i g_{i}^{-1} $ between the stabilizer subgroups of the deformations.\footnote{Due to this, we can drop the stabilizer subgroups from our notation, but we write them explicitly whenever they are required.} The horizontal composition of two $2$-morphisms $\alpha_1$ and $\beta_1$ is naturally defined as follows. Let
\beal
\alpha_1 &=\{(g_1 H_1, {\cal O}_1),\ldots,\, (g_n H_n, {\cal O}_n)\,\}\,, \\ 
\beta_1 &=\{(k_1 L_1, {\cal U}_1),\ldots,\, (k_m L_m, {\cal U}_m)\,\}\,,
\eeal
the horizontal composition then is\footnote{Note that the horizontal composition of $\alpha_1$ and $\beta_1$ requires the target of $\{{\cal O}_i\}_{i=1}^n: X_1\rightarrow X_{n+1}$ match the source of $\{{\cal U}_\ell\}_{\ell=1}^m:X'_1 \rightarrow X'_{m+1}$, {\it i.e.} $X_{n+1}=X'_1$.}
\beal
\beta_1 \circ \alpha_1 = \; & \{(g_1, {\cal O}_1),\, \ldots, \, (g_n, {\cal O}_n), \\
& \qquad\qquad \,  (k_1, {\cal U}_1),\,\ldots,\, (k_m, {\cal U}_m),\}
\eeal
as we  concatenate two sequences of deformation. %
 This implies that any $2$-morphism of the form \eqref{eq: 2-morph} can be expressed as the horizontal composition of a sequence of {\it basic} $2$-morphisms $\alpha_i = (g_i, {\cal O}_i)$, which act on basic deformations. %
 %
%
Let us now verify that the interchange law is satisfied (see Figure \ref{inter}). First,\footnote{Here the $2$-morphism $\beta_2$ is defined as the sequence $ \{ (p_1,{\cal U'}_1),\, \ldots, \, (p_m, {\cal U'}_m) \}$.}
\be
&& (\beta_2 \bullet \beta_1) \circ (\alpha_2 \bullet  \alpha_1) = \\
&&  \qquad = \{(p_1  k_1, {\cal U}_1),\,\ldots,\, (p_k  h_k, {\cal U}_m)\} \circ \nonumber\\
&& \qquad\qquad\quad \, \circ \,  \{(h_1 g_1 , {\cal O}_1),\,\ldots,\, (h_n g_n , {\cal O}_n)\} \nonumber \\
&& \qquad = \{(h_1  g_1, {\cal O}_1),\, \ldots,  (h_n  g_n, {\cal O}_n),\, \nonumber \\
&& \qquad\qquad\qquad \, (p_1  k_1, {\cal U}_1),\, \ldots, \, (p_m  k_m, {\cal U}_m)\}\,. \nonumber
\ee
Whereas,
\be
&& \hspace{-1em} (\beta_2\circ \alpha_2) \bullet (\beta_1\circ \alpha_1) =\\
&& = \{(h_1, {\cal O'}_1),\, \ldots, \, (h_n, {\cal O'}_n),\, (p_1, {\cal U'}_1),\, \ldots,\, (p_m, {\cal U'}_m)\} \bullet \nonumber \\
&& \quad\quad \bullet \, \{(g_1, {\cal O}_1),\, \ldots, \, (g_n, {\cal O}_n), (k_1, {\cal U}_1),\, \ldots, \, (k_m, {\cal U}_m)\}\nonumber\\
&& = \{(h_1  g_1, {\cal O}_1),\, \ldots,  (h_n  g_n, {\cal O}_n),\, \nonumber \\
&& \qquad\qquad \, (p_1  k_1, {\cal U}_1),\, \ldots, \, (p_m  k_m, {\cal U}_m)\}\,. \nonumber
\ee
We thus see that,
\begin{equation}
(\beta_2 \bullet \beta_1) \circ (\alpha_2 \bullet  \alpha_1)  =
(\beta_2\circ \alpha_2) \bullet (\beta_1\circ \alpha_1)\,,
\end{equation}
and the interchange law holds true. The category of CFTs together with the action of the $0$-form symmetry thus forms a strict $2$-category structure. 

Note that gauging can be incorporated in the same structure. We take a subgroup of $G^{(X)}_0$ for a CFT $X$ and gauge it. This breaks $G^{(X)}_0$ to the centralizer of the gauge subgroup and further removes anomalous abelian factors. To gauge a symmetry we choose an embedding of the gauge group in $G^{(X)}_0$. We thus can discuss different equivalent ways to do so which are related by an action of the global symmetry leading to $2$-morphisms.  Similarly we can discuss irrelevant deformations being morphism from an object to itself and exactly marginal deformations which take us between different objects but with no flow involved. We will discuss the latter next.

\section{Conformal manifold $2$-category}
Let us consider the special case of exactly marginal deformations. 
Most of the concrete examples of theories with exactly marginal operators involve supersymmetric CFTs,\footnote{See {\it e.g.} \cite{Leigh:1995ep,Green:2010da,Komargodski:2020ved,Razamat:2020pra,Gomis:2015yaa,Baggio:2017mas,Baggio:2017aww,Niarchos:2021iax,Perlmutter:2020buo} for some results in the supersymmetric case.} though special degenerate cases of theories without supersymmetry are known to exist. See {\it e.g.} the discussion in \cite{Bashmakov:2017rko}. %
Exactly marginal deformations of supersymmetric theories parametrize what is called the conformal manifold ${\cal M}_c$.
Here no RG flow is triggered on ${\cal M}_c$: the values of the couplings are essential as they determine the target CFT and have a geometrical meaning  as (local) coordinates on the conformal manifold. %
As a result, our procedure of defining $1$-morphisms with a sequence of flows better be interpreted as a concatenation of infinitesimal exactly marginal deformations. Geometrically, each such concatenation consists of a series of small consecutive deviations from the source CFT point along the conformal manifold: it forms a path in ${\cal M}_c$. Therefore, we conclude that $1$-morphisms between CFTs on the same conformal manifold are paths between the corresponding points, such that different paths correspond to different morphisms.\footnote{
Note that here it is somewhat natural to identify homotopically equivalent paths. In that case, every morphism becomes invertible, with the inverse given by the oppositely oriented path. This gives the category of the CFTs residing on the same conformal manifold the structure of the {\it path groupoid} \, ${\mathbb P}_1({\cal M}_c)$ of the conformal manifold
 (if we identify homotopically equivalent deformations). As we will soon discuss, in some cases one can associate more than one 1-morphism to a given path. We can consider the skeletal category of the conformal manifold groupoid. If we identify morphisms corresponding to the same path this is  given by the homotopy group of the conformal manifold. Moreover, in the skeletal category of  ${\cal C}^{(D)}$ theories residing on the same conformal manifolds will be identified as objects.}

\begin{figure}
\includegraphics[scale=0.2]{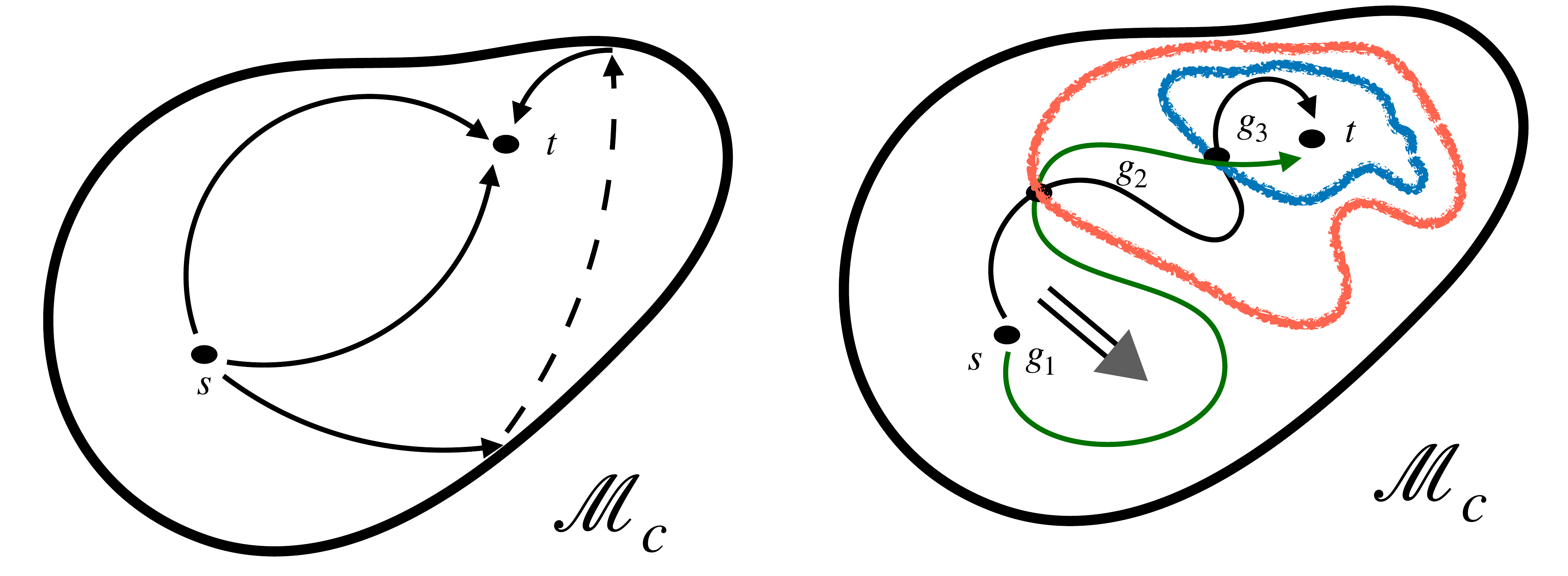}
\caption{ \label{confm}A conformal manifold. On the left we depict morphisms corresponding to paths. On the right, the shaded blue and orange lines denote loci with enhanced symmetry. The two $1$-morphisms (black and green paths) are related by a $2$-morphism $\alpha=\{(g_1, {\cal O}_1), (g_2, {\cal O}_2), (g_3, {\cal O}_3)\}$, defined via the intersection of the paths (corresponding to the $1$-morphisms) and the enhancement loci. In general, when leaving one such locus we have multiple choices of how to break the symmetry, all related to each other by the action of elements of the enhanced symmetry group. These group elements $g_i$ determine the structure of $\alpha$.}
\end{figure}

Turning to $2$-morphisms, we should look at the global symmetry. On generic points of the conformal manifold, it is described by the group $G_{{\cal M}_c}$. However, there might be special loci within ${\cal M}_c$ where the symmetry gets enhanced to a bigger group $G_\text{locus} \supset G_{{\cal M}_c}$. 
By definition, all the exactly marginal deformations preserve  $G_{{\cal M}_c}$, but on an enhancement locus, where the symmetry is larger, various deformations might be again related to each other by the action of the enhanced global symmetry group. This gives us the $2$-category structure as before. Here however, each $2$-morphism is parametrized by a sequence of (equivalence classes of) group elements  associated to loci where the two paths in question intersect a locus with enhanced global symmetry.\footnote{To be precise, here the $2$-morphisms are parametrized by pairs, where the first element belongs to the appropriate coset  and the second element is the corresponding source deformation.} Therefore, the source and target $1$-morphisms might correspond to the same path or different paths which intersect at least at the same loci of enhanced symmetry (see Figure \ref{confm}). 
The fact that the
 loci with enhanced symmetry behave differently than the generic one is because these loci will have more marginal operators (which are marginally irrelevant in supersymmetric theories) \cite{Green:2010da}. Thus, turning on generic marginal deformations, we do generate a flow which ends up on the same conformal manifold. 
 The $2$-category structure  is intimately related with the symmetry structure of the conformal manifold.
 
\section{Monoidal structure}

The category ${\cal C}^{(D)}$ we have built admits a natural monoidal structure, {\it i.e.}~a tensor product. Given $X,Y\in {\cal C}^{(D)}_0$, we obtain a new object $Z \equiv X \otimes Y \in {\cal C}^{(D)}_0$ by taking the tensor product of Hilbert spaces and (extended) operator algebras 
of the  theories $X$ and $Y$. This produces the decoupled sum of the original degrees of freedom and thus it forms a consistent theory. If $X$ and $Y$ both admit a Lagrangian description, the path integral of $Z$ is nothing but the product of the path integrals of $X$ and $Y$ (the action of $Z$ is simply the sum $S_Z = S_X + S_Y$), therefore all correlation functions factorize. The {\it unit object} $\Id$ is given by the empty theory, with no dynamical fields and vanishing action, such that its product with any other object leaves the latter invariant. Under these assumptions, the tensor product is manifestly associative. 

In our construction, fields of particular spin, namely the scalar field and the fermions of various types in a given dimension, are {\it elementary} objects  in the sense that they are not tensor products of other objects. Theories with Lagrangian description are constructed by tensoring these elementary objects and following morphisms. The free matter fields are not necessarily the only elementary objects (see Section \ref{sec: discussion} for more details). Note that spin one fields in our discussion play a different role compared to other spins as these are associated to morphisms.

The tensor product admits a natural lift to a tensor product of $1$- and $2$-morphisms, which makes it into a $2$-functor on the $2$-category ${\cal C}^{(D)}$. Namely, given two source CFTs $X$ and $Y$ and the sequences of deformations $f : X \to X'$ and $\textsl{g} : Y \to Y'$, we can define the 1-morphism from $X\otimes Y$ to $X' \otimes Y'$, which we will denote by $f \otimes \textsl{g}$, as the combination of the two sequences of deformations. This operation preserves compositions and identity 1-morphisms. Similarly, given the deformations $f, f' : X \to X'$ and $\textsl{g}, \textsl{g}' : Y \to Y'$, such that there exist $\alpha : f \Rightarrow f'$ and $\beta : \textsl{g} \Rightarrow \textsl{g}'$ defined as above, we can build the $2$-morphism $\alpha \otimes \beta : f \otimes f' \Rightarrow \cg \otimes \cg'$ by combining the sequences describing $\alpha$ and $\beta$. Once again, this operation preserves compositions and identity $2$-morphisms. The structure that we obtain is that of a strict monoidal $2$-category $({\cal C}^{(D)}, \otimes)$. 

\begin{figure}
    \centering
    \includegraphics[scale=0.13,trim={5px 5px 5px 5px},clip]{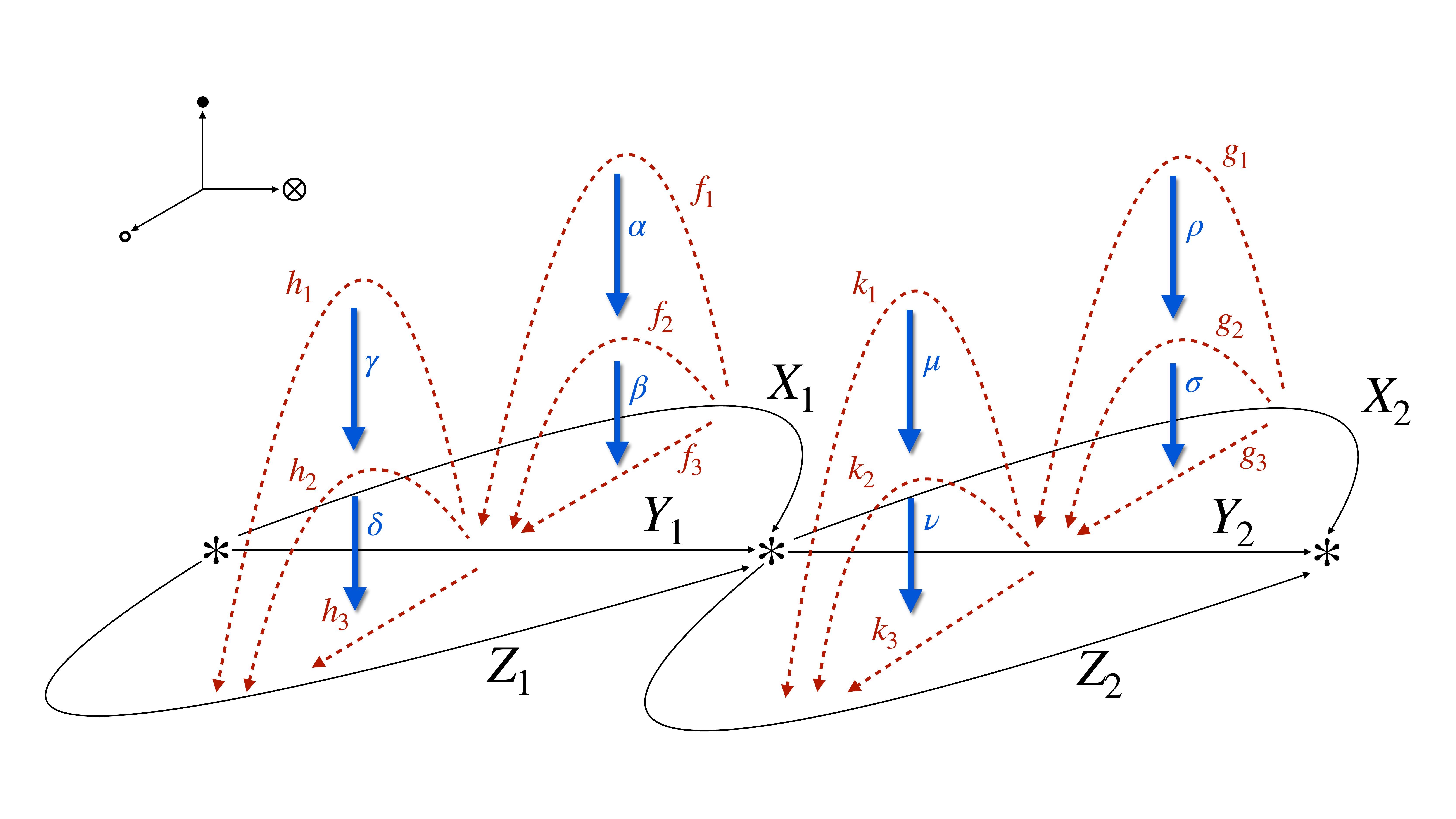}
    \caption{Illustration of the interchange laws. We have a single object $*$ in the $3$-category and the $1$-morphisms are the CFTs.
    Each axis corresponds to a different morphism: the $x$ axis is the tensor product $\otimes$, the $y$ axis is the composition $\circ$ of the category ${\cal C}^{(D)}$, while the $z$-axis is the vertical product $\bullet$ of the $2$-morphisms of ${\cal C}^{(D)}$.}
    \label{fig:3cat}
\end{figure}

A natural way of understanding the structure of ${\cal C}^{(D)}$ is to view it as a (strict) 3-category, say ${\cal D}^{(D)}$, with a single object denoted by $*$ \cite{MR1712872}. Here the $1$-morphisms of ${\cal D}^{(D)}$ are the objects of ${\cal C}^{(D)}$, viewed abstractly as endomorphims of $*$. Thus, the $2$- and $3$-morphisms of ${\cal D}^{(D)}$ are nothing but the $1$- and $2$-morphisms of ${\cal C}^{(D)}$ respectively. 
The three compositions on ${\cal D}^{(D)}$ are then constructed as follows (see Figure \ref{fig:3cat}). The vertical composition of $3$-morphisms of ${\cal D}^{(D)}$ coincides with the $\bullet$ of ${\cal C}^{(D)}$. Next, the functor describing the horizontal composition of the $2$-morphisms of ${\cal D}^{(D)}$, and its lift to $3$-morphisms, is the horizontal composition $\circ$ of ${\cal C}^{(D)}$. Finally, the 2-functor providing the composition of $1$-morphisms of ${\cal D}^{(D)}$, together with its lift to $2$- and $3$-morphisms, is the tensor product $\otimes$.
The axioms of a $3$-category require that these three different structures  are compatible with each other and satisfy
some properties, the interchange laws:
\be\label{intlaws}
&&\left(\delta\bullet\gamma\right)\circ\left(\beta\bullet\alpha\right)  =\left(\delta\circ\beta\right)\bullet\left(\gamma\circ\alpha\right)\,,\nonumber\\
&&\left(\sigma\otimes\beta\right)\bullet\left(\rho\otimes\alpha\right)  =\left(\sigma\bullet\rho\right)\otimes\left(\beta\bullet\alpha\right)\,,\\
&&\left(\mu\otimes\gamma\right)\circ\left(\rho\otimes\alpha\right)  =\left(\mu\circ\rho\right)\otimes\left(\gamma\circ\alpha\right)\,.\nonumber
\ee
These can be read from the three planes in Figure \ref{fig:3cat}.

The first interchange law is the one we already checked above for $2$-morphisms
in $\mathcal{C}^{(D)}$. To prove the others, we use the definitions for $1$-morphisms
and $2$-morphisms in $\mathcal{C}$. We denote,
\be 
&&f \,: X_1\to Y_1\,   = \left\{ \mathcal{O}_{1},\dots,\mathcal{O}_{n}\right\} \,,\\
&&\;\textsl{g} \,: X_2\to Y_2\,  =\left\{ \mathcal{U}_{1},\dots,\mathcal{U}_{m}\right\}\,.\nonumber 
\ee
Then, we write their tensor product as
\be
&&f\otimes \textsl{g}\,: X_1\otimes X_2\to Y_1\otimes Y_2\, =\\
&&\qquad\qquad\qquad \left\{ \mathcal{O}_{1},\cdots,\mathcal{O}_{n},\mathcal{U}_{1},\cdots,\mathcal{U}_{m}\right\}\,. \nonumber
\ee Note that here the ordering between the sets of $\mathcal{O}$'s and $\mathcal{U}$'s does not matter. Only the relative ordering between different ${\cal O}$'s (resp.~${\cal U}$) does. Therefore the product is commutative: $f\otimes \textsl{g}=\textsl{g}\otimes f$.
Next we employ the following notation for $2$-morphisms:
\be 
\alpha=\left\{ \left(g_{1}^{(\alpha)},\mathcal{O}_{1}\right),...,\left(g_{n}^{(\alpha)},\mathcal{O}_{n}\right)\right\} \, ,
\ee 
so that the tensor product of $2$-morphisms reads,\footnote{Here we consider the tensor product of the groups acting on the two tensored components. Note that the global symmetry of the tensor product of two theories might be bigger than the tensor product of the symmetries of the two components. An example is tensoring a collection say of $2$ complex scalar fields. However, for the purpose of checking the interchange laws this enhancement is not essential.}
\be 
&&\rho\otimes\alpha=\left\{ \left(\mathds{1}\otimes g_{1}^{(\rho)},\mathcal{U}_{1}^{\textsl{g}_1}\right),\dots,\right.\\
&&\left.\left(\mathds{1}\otimes g_{m}^{(\rho)},\mathcal{U}_{m}^{\textsl{g}_1}\right), \left(g_{1}^{(\alpha)}\otimes \mathds{1},\mathcal{O}^{f_1}_{1}\right),\dots,\left(g_{n}^{(\alpha)}\otimes \mathds{1},\mathcal{O}_{n}^{f_1}\right)\right\} \,. \nonumber 
\ee  
Using the definitions of compositions we then can prove the interchange laws \eqref{intlaws}. For example,
\begin{widetext}
\be 
&&\left(\sigma\otimes\beta\right)  \bullet\left(\rho\otimes\alpha\right)=
\left\{ \left(\mathds{1}\otimes g_{1}^{(\sigma)},\mathcal{U}_{1}^{\textsl{g}_2}\right),\cdots ,\left(\mathds{1}\otimes g_{m}^{(\sigma)},\mathcal{U}_{m}^{\textsl{g}_2}\right), \left(g_{1}^{(\beta)}\otimes \mathds{1},\mathcal{O}_{1}^{f_2}\right),\cdots ,\left(g_{n}^{(\beta)}\otimes \mathds{1},\mathcal{O}_{n}^{f_2}\right)\right\} 
  \bullet\\&& \quad \bullet \left\{ \left(\mathds{1}\otimes g_{1}^{(\rho)},\mathcal{U}_{1}^{\textsl{g}_1}\right),\dots ,\left(\mathds{1}\otimes g_{m}^{(\rho)},\mathcal{U}_{m}^{\textsl{g}_1}\right), \left(g_{1}^{(\alpha)}\otimes \mathds{1},\mathcal{O}_{1}^{f_1}\right),\dots ,\left(g_{n}^{(\alpha)}\otimes \mathds{1},\mathcal{O}_{n}^{f_1}\right)\right\}\nonumber\\&&
=  \left\{ \left( \mathds{1}\otimes g_{1}^{(\sigma)}g_{1}^{(\rho)},\mathcal{U}_{1}^{\textsl{g}_1}\right),\dots ,\left(\mathds{1}\otimes g_{m}^{(\sigma)}g_{m}^{(\rho)},\mathcal{U}_{m}^{\textsl{g}_1}\right), \left(g_{1}^{(\beta)}g_{1}^{(\alpha)}\otimes \mathds{1},\mathcal{O}_{1}^{f_1}\right),\dots,\left(g_{n}^{(\beta)}g_{n}^{(\alpha)}\otimes \mathds{1}, \mathcal{O}_{n}^{f_1}\right)\right\} \nonumber\\&&
=  \left(\left\{ \left(g_{1}^{(\sigma)},\mathcal{U}_{1}^{\textsl{g}_2}\right),\dots ,\left(g_{m}^{(\sigma)},\mathcal{U}_{m}^{\textsl{g}_2}\right)\right\} \bullet\left\{ \left(g_{1}^{(\rho)},\mathcal{U}_{1}^{\textsl{g}_1}\right),\dots,\left(g_{m}^{(\rho)},\mathcal{U}_{m}^{\textsl{g}_1}\right)\right\} \right) \otimes \nonumber\\&& \quad
  \left(\left\{ \left(g_{1}^{(\beta)},\mathcal{O}_{1}^{f_2}\right),\dots ,\left(g_{n}^{(\beta)},\mathcal{O}_{n}^{f_2}\right)\right\} \bullet\left\{ \left(g_{1}^{(\alpha)},\mathcal{O}_{1}^{f_1}\right),\dots ,\left(g_{n}^{(\alpha)},\mathcal{O}_{n}^{f_1}\right)\right\}\right)\nonumber\\&&
=  \left(\sigma\bullet\rho\right)\otimes\left(\beta\bullet\alpha\right)\,.\nonumber
\ee 
\end{widetext}
In this equation, the superscripts of ${\cal O}$ and ${\cal U}$ denote the $1$-morphisms to which the basic deformations belong.
Also, as usual, $\mathds{1}$ here denotes the identity element of a corresponding group.

\section{Category of CFTs in any dimension}

Let us next define the category of all CFTs ${\cal C}$. The set of objects in ${\cal C}$ is given by ${\cal C}_0=\cup_{D=0}^\infty {\cal C}^{(D)}_0$.
The set of morphisms includes the morphisms of ${\cal C}^{(D)}$,
\begin{equation}
\cup_{D=0}^\infty {\cal C}^{(D)}_1\subset {\cal C}_1\,,
\end{equation} but also contains additional morphisms connecting objects in ${\cal C}^{(D)}_0$ with different values of $D$. Physically we define these additional morphisms as follows. Given $X \in {\cal C}^{(D)}_0$ we can place it on a  $M^{(D')} \times m^{(D-D')}$ where $m^{(D-D')}$ is a compact space. We can also turn on background gauge fields supported on $m^{(D-D')}$, $\{{\bf {\cal A}}\}$. Then we can discuss the low energy physics of this construction. The resulting physics might be describable by a $D'$-dimensional CFT $Y \in {\cal C}^{(D')}_0$. If that is the case we define a morphism $X \to Y$ which is parametrized by the compactification geometry $\{m^{(D-D')},\, {\bf{\cal A}\}}$ and the source CFT. We will refer to these as {\it across-dimensions} morphisms.

As was the case with {\it in-dimension} morphisms, we can naturally compose various across-dimension morphisms with each other and also across-dimension morphisms with in-dimension morphisms. A general morphism is thus defined by an ordered sequence of deformations, some of which are in-dimension deformations and some are across-dimension compactifications.

Next, the compactification deformation might preserve the $0$-form symmetry of the source CFT or it might break it, say by the choice of the background gauge fields.  Different compactifications preserving the same symmetry of the source CFT thus again might be related by the action of the $0$-form symmetry group of the source CFT. This provides $2$-morphisms between compactification $1$-morphisms. More generally, a morphism defined by a sequence of deformations can be related by a sequence (of equivalence classes) of group elements as before. This provides a $2$-category structure on  ${\cal C}$.  Note that, abstractly, this structure does not distinguish in-dimension and across-dimension morphisms and treats them uniformly.   

The $0$-form symmetry of the lower dimensional target theory can have several higher dimensional origins. First, it can be the $0$-form symmetry of the source CFT. Second, it can come from definitions of boundaries of the compactification geometries. Finally, it can come from higher form symmetry of the source CFTs taking the corresponding topological operators to wrap the compactification surface.
In addition, some of the $0$-form symmetry of the target CFT might be emergent as usual. Finally, also non-topological operators, local and non-local operators can give rise to local operators in lower dimensions which we can associate to morphisms, see {\it e.g.}~\cite{babuip} and Appendix E of \cite{Kim:2017toz}.

To define a monoidal structure, we need to extend the set of objects to incorporate coupled CFTs of different dimensionalities. One can naturally consider a product of two CFTs of different dimensionalities $D$ and $D'<D$ by considering some $D'$ dimensional hyperplane in $D$ dimensional space and placing the $D'$-dimensional CFT on it. The bulk CFT and the lower dimensional one are not coupled. Next, one can consider coupling the two CFTs in various ways, {\it e.g.}~gauging symmetries or coupling operators. If the resulting theory is conformally invariant we can add it to the set of objects. Note that tensoring more than two objects requires a more rigorous definition: {\it e.g.}~one can define tensored objects of same dimensionality to share the same spacetime, and objects of lower dimensionality to live on submanifolds of objects all of higher dimensionality.
The resulting category we will denote by $\widetilde {\cal C}$. The category $\widetilde {\cal C}$ can be thought of as category of CFTs in arbitrary number of dimensions in presence of arbitrary conformal defects. 

\section{Smaller categories}

One can discuss various ways to simplify or constrain the category of all CFTs. One way to do so  has been already discussed: we can consider the set of theories residing on the same conformal manifold. This restriction retains the structure of $2$-category but does not have a natural tensor product.
Another way to obtain a more general class of theories is to consider CFTs with at least some amount of supersymmetry. 
For example, one can define a category ${\cal C}^{(D=4|{\cal N}=1)}$ such that  ${\cal C}^{(D=4|{\cal N}=1)}_0$ is the subset of   ${\cal C}^{(D=4)}_0$ corresponding to theories which have at least $D=4$ ${\cal N}=1$ superconformal symmetry (are SCFTs). To define morphisms between the different theories here we need to be more careful. For example, if we want the deformations to preserve supersymmetry explicitly we need to turn on first several deformations at once, scalar potentials and Yukawa terms following from a superpotential. Second, gauging a symmetry preserving ${\cal N}=1$ supersymmetry corresponds to adding not just vector fields but also gaugino fermions. In our general definition this gauging, thus, corresponds to tensoring with free fields, gauging, and turning on potentials. One can define morphisms using these supersymmetric definitions and hence consider the category of the supersymmetric theories with morphisms being supersymmetric deformations.\footnote{Note that with the supersymmetric definition of morphisms we can have relevant deformations starting and ending on the same object. A simple example is the ${\cal N}=2$ duality in $D=3$ between a single chiral superfield and $U(1)$ gauge theory with a single charge $+1$ chiral superfield \cite{Dimofte:2011ju}.} This construction can be generalized to a category of supersymmetric CFTs in any dimensions. Here one would insist discussing compactifications between dimensions preserving some amount of supersymmetry. This implies that one has in general to consider {\it twisting} along with compactification: that is turning nontrivial background fields for the R-symmetry. The $2$-category structure works in the same way as before.

 \begin{figure}
\includegraphics[scale=0.18]{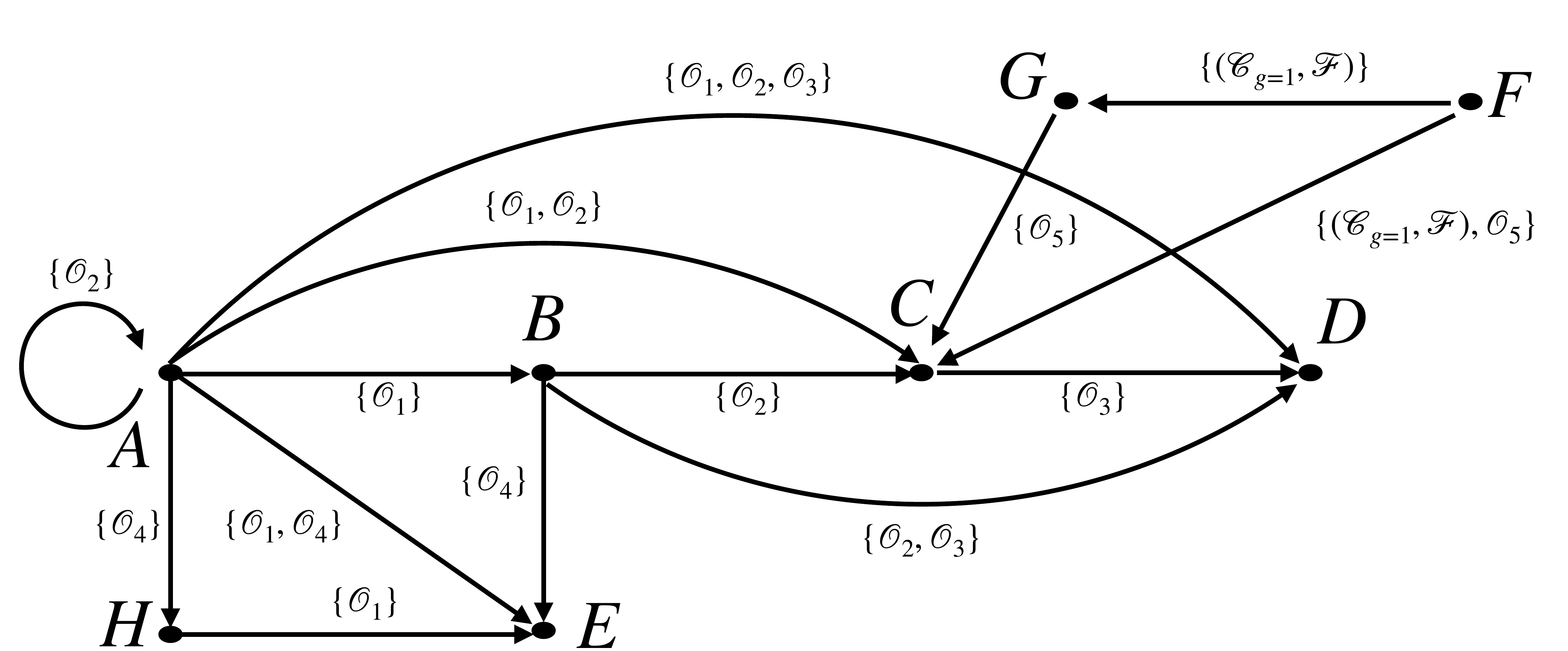}
\caption{ \label{SU2flows} An example of a network of flows. The source theory $A$ is a collection of thirty one chiral superfields (complex scalars and Weyl fermions). The various morphisms and other objects are discussed in the text. The  lines  $\{{\cal O}_1,\,{\cal O}_2,\,{\cal O}_3\}$ and $\{{\cal O}_2,\,{\cal O}_3\}$  correspond to morphisms which are associated to deformations which can be only defined as a sequence but cannot be thought of directly as a  deformation of the source theory. }
\end{figure}

\section{Some examples}

\noindent{{\it Example I}: }\;\; Let us discuss an example of objects and morphisms in a category of supersymmetric CFTs: all of the morphisms and objects will preserve some amount of supersymmetry. We can phrase the sequence of flows in non-supersymmetric category but this would be more cumbersome. Consider as the source CFT $A$ the collection of $15+16=31$ ${\cal N}=1$ chiral superfields in four dimensions. These include complex scalars and Weyl fermions. Next we turn on deformation $\{{\cal O}_1\}$ which corresponds to splitting the fields into $2\times 8+15$ and gauging $SU(2)$ subgroup of the $U(31)$ global (non R) symmetry of $A$. Here we consider the symmetries preserving the structure of supermultiplets. The fields form $15$ singlets and $8$ fundamentals. This is a relevant deformation which takes us to CFT $B$. We can then, for example, take two of the eight doublets and form from them a mesonic operator and deform $B$ by turning on a superpotential for this operator. This is a relevant deformation $\{{\cal O}_4\}$. In the IR this flows to CFT  $E$. Note that from point of view of $A$ we turned on a mass term and thus the theory $E$ is $SU(2)$ SQCD with $N_f=3$ and additional $15$ free chiral fields \cite{Seiberg:1994pq}. This SQCD flows in IR to $15$ free fields and thus  CFT $E$ is a collection of $15+15=30$ free fields. We can consider this deformation directly at   $A$ and then we label it as $\{ {\cal O}_1, \, {\cal O}_4\}$. Alternatively, we could first turn on the mass term which would lead to free CFT of $12+15=29$ fields in the IR, $H$, and then gauge $SU(2)$ group with now six fundamental fields, leading again to $E$. Let us consider now starting from $B$. This theory has $SU(8)\times U(15)$ global symmetry and $28$ operators in ${\bf 28}$ of $SU(8)$  which can be thought as mesons and baryons of $A$ after gauging. Let us consider an $SU(2)\times SU(6)\times U(1)$ subgroup of $SU(8)$ under which ${\bf 28}\to ({\bf 1},{\bf 15})\oplus({\bf 2},{\bf 6})\oplus ({\bf 1},{\bf 1})$. We couple the $ ({\bf 1},{\bf 15})$ to the fifteen free fields in the superpotential and denote this deformation by $\{{\cal O}_2\}$. This is a relevant deformation leading to CFT $C$.
 The theory $C$ has (conjecturally) an emergent symmetry $SU(2)\times SU(6)\to E_6$ \cite{Razamat:2017wsk}. Note that, had we performed the deformation $\{{\cal O}_2\}$ directly on $A$, we would have obtained a cubic superpotential in the free theory, which is an irrelevant deformation leading us back to $A$. Hence the deformation $\{{\cal O}_1,\,{\cal O}_2\}$ represents an example of a {\it dangerously irrelevant} deformation on $A$ when intended as turning on both ${\cal O}_1$ and ${\cal O}_2$ simultaneously: however, as we stressed above, the deformations are considered to be taken one by one in a sequence of specified order.  Here,  $$\{{\cal O}_1,\,{\cal O}_2\} \neq \{{\cal O}_2,\,{\cal O}_1\}\sim\{{\cal O}_1\}.$$ The equivalence on the right just means that the target CFTs are the same but we will distinguish the two morphisms.
 Next,
  the $E_6$ global symmetry has $SU(3)^3$ maximal subgroup with one of the three $SU(3)$ factors emerging in the IR.
  We can then consider various deformations making use of the emergent symmetry. For example, we can compactify the theory on a circle to three dimensions and gauge a diagonal combination of the three $SU(3)$s in the IR turning on a Chern-Simons term with some level:  we denote this deformation by $\{{\cal O}_3\}$ which is by iteself a concatination of two deformations (compactification and gauging). 
  This leads to CFT $D$. Note that we can consider the deformation $\{{\cal O}_1,\,{\cal O}_2,\, {\cal O}_3\}$ starting from $A$ and leading to $D$. 
  However, this deformation cannot be defined field theoretically  in $A$ as we gauge an emergent symmetry and only makes sense as a sequence of deformations. Finally we can start from an SCFT in six dimension---the rank one E-string theory \cite{Ganor:1996mu,Seiberg:1996vs,Witten:1996qb,Morrison:1996pp} --- which we denote by $F$.
  This theory has $E_8$ global symmetry. We can then deform it by placing it on a torus with a flux breaking $E_8$ to $E_6\times U(1)$. We denote this deformation as  $\{({\cal C}_{g=1},\, {\cal F})\}$. The theory will flow to a four dimensional CFT, $G$. A relevant superpotential deformation of $G$, denoted by $\{{\cal O}_5\}$ leads again to $D$. See \cite{Razamat:2017wsk,Razamat:2022gpm} for details.
  
We have discussed here some flows starting from $A$ and $F$: the resulting objects and morphisms are part of a much larger categorical structure and we only used the above as an illustration. Note that at each step we had a choice of a given subgroup to define the deformation. Different choices  lead to equivalent theories in the IR, and thus the relevant deformations are related by $2$-morphisms defined by mapping one choice into the other one.

\

 \begin{figure}
\includegraphics[scale=0.3]{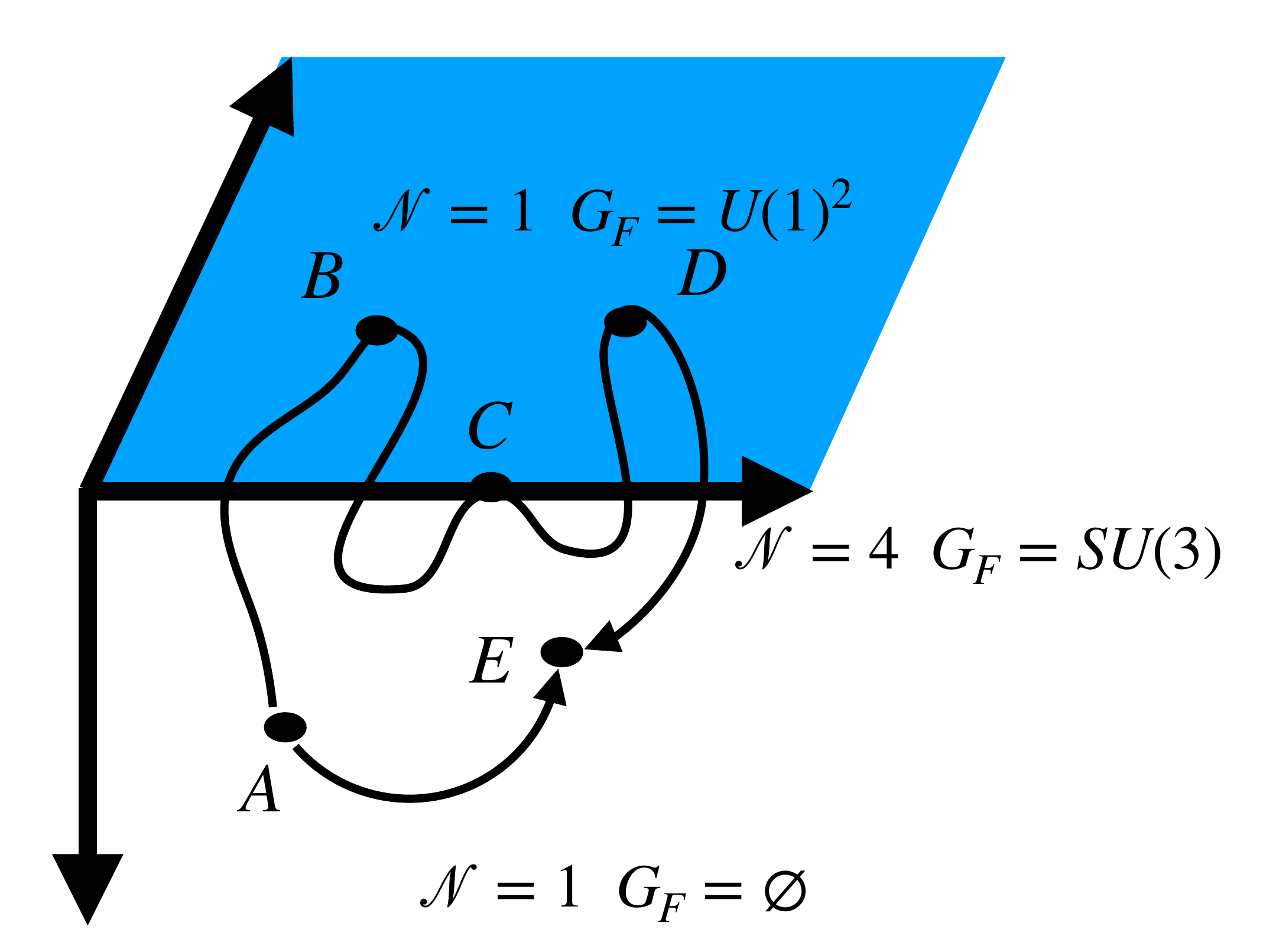}
\caption{ \label{N4Mc} Conformal manifold of ${\cal N}=4$ $SU(N>2)$ SYM. The manifold has three complex dimensions. Along one of the dimensions the supersymmetry is ${\cal N}=4$ and the global symmetry in ${\cal N}=1$ language is $SU(3)$. Along two dimensions the supersymmetry is ${\cal N}=1$ and symmetry is $U(1)^2$ generically. While on general locus supersymmetry is ${\cal N}=1$ and there is no continuous global symmetry.}
\end{figure}

\noindent{{\it Example II}: }\;\; Next, let us consider the conformal manifold of ${\cal N}=4$ SYM with $SU(N)$ gauge group. For $SU(N>2)$ the conformal manifold has three complex dimensions. Along one of the complex directions the supersymmetry is ${\cal N}=4$. Viewing this theory as an ${\cal N}=1$ SCFT, the global symmetry along this direction is $SU(3)$. Along the two additional complex directions the supersymmetry is broken to ${\cal N}=1$. One of these direction preserves a $U(1)^2$ subgroup of $SU(3)$ while on a generic locus of the conformal manifold the continuous global symmetry is completely broken and one only has R-symmetry and supersymmetry \cite{Leigh:1995ep}. General $1$-morphisms between two CFTs on the manifold correspond to continuous paths. If  the path passes through locus of enhanced symmetry, one has a choice of embedding for the deformation that  breaks the enhanced global symmetry group once the path leaves the enhanced locus. If we have 
two paths between the same pairs of points on ${\cal M}_c$ which pass through same loci of enhanced symmetry we can define a $2$-morphism between  them. The $2$-morphism is parametrized by a sequence of (equivalence classes of) group elements of the enhanced symmetry transforming the choice of one deformation into the other.

\section{Discussion and Comments}
\label{sec: discussion}

In this note, we have discussed a categorical language to organize our thinking of the space of CFTs.
This discussion fits the general framework of higher form/higher group/categorical symmetries. The layer we are trying to add corresponds to deformations of a CFT. The higher category of generalized symmetries
associated to a given theory acts on various operators in that theory. In particular, some of these operators can be used to deform a given CFT to a new CFT. These deformations are  $1$-morphisms of a category, while the  
symmetries provide a higher categorical structure. Another type of morphisms is given by gauging  some (generalized) symmetries. In particular, we have discussed in detail the application of this idea to operator deformations and gauging of $0$-form symmetries.\footnote{
The deformations can be thought of as space-time filling objects. For example, the operator deformations are terms in the action. These deformations are not topological in the usual sense and thus do not correspond to what is often called $(-1)$-form symmetries \cite{Cordova:2019uob,Vandermeulen:2022edk}.
However, as we are interested only in the fixed points, the fine details of the values of relevant and irrelevant couplings are inessential and one can view this as a topological property.}
There are various 
ways in which the discussion can be extended. For example, we can consider gauging higher form/group symmetries \cite{Tachikawa:2017gyf}. The gauging of such symmetries does not lead to RG flow but does change the spectrum of operators of different dimensionality the theory has and thus leads to a different CFT. Moreover one can also consider gauging global symmetries of various forms on submanifolds of various codimensions \cite{Roumpedakis:2022aik}.

As our main motivation to develop the categorical language is to discuss various conjectures and questions regarding the space of all CFTs (with the hope that such a reformulation will eventually lead to deeper insights),  
let us list some of the questions/conjectures.\footnote{See also \cite{razamatstrings22}.}
\begin{itemize}
\item {\it Is there a morphism in ${\cal C}^{(D)}_1$ to any given CFT from an object corresponding to a free theory in $D\leq 4$?}  Remember that a free CFT is a tensor product of some number of free scalars and free fermions. This question amounts to wondering whether any CFT has a Lagrangian construction in a given number of dimensions. Note that by Lagrangian construction here we include sequences of deformations.
We can phrase this as asking whether one can define a set of {\it elementary objects} (which might not be unique) such that: (i) it includes free matter theories (ii)  all the other theories are obtained from it by tensor products and deformations; and whether this set of elementary theories is strictly larger than the set of free theories.  This question can be refined in various ways. 
%

\item  {\it Is there a supersymmetric morphism  to any given SCFT from an object corresponding to a free theory in $D\leq 4$?} This question might be refined by demanding the deformations and collections of free fields to be also supersymmetric.

\item{\it Is there a morphism in ${\cal C}_1$ to any given CFT starting from an object corresponding to a CFT in $D=6$?} Here we wonder whether any CFT in lower dimensions can be obtained as a compactification, and possibly subsequent deformation, of a six dimensional CFT.

\item{\it Is any $D\leq 4$  (S)CFT obtained from a six dimensional CFTs also in the target of free CFTs?} That is, whether all compactifications are across-dimensions dual to lower dimensional field theoretic constructions.

\item {\it What are the  nontrivial objects with no outgoing morphisms which are not TQFTs?} Such theories are sometimes called dead-end CFTs \cite{Nakayama:2015bwa,Frenkel:1988xz}.

\item Studying the structure of theory space led in the past to various explicit quantitative results. An example is the relation between compactifications of $6d$ CFTs on surfaces and supersymmetric partition functions \cite{Pestun:2016zxk} of lower dimensional theories. 
Here, the supersymmetric partition functions can be either labeled by the target lower dimensional CFT, {\it e.g.} $Z[T_{4d}]$, and then typically  hard to compute, or by the across-dimensions morphisns, {\it e.g.} $Z[T_{4d}]=Z[\left(m^{(2)},\{{\cal A}\}\right),T_{6d})]$, and then often easier to derive. See {\it e.g.} \cite{Alday:2009aq,Gadde:2009kb}.
\end{itemize}

\

\noindent{\bf Acknowledgments}:~
We are grateful to Chris Beem, Dan Freed, Zohar Komargodski, Elli Pomoni, Sakura Schafer-Nameki, Yuji Tachikawa, and Amos Yarom  for insightful discussions and comments.
This research is supported in part by Israel Science Foundation under grant no. 2289/18, grant no. 2159/22, by I-CORE  Program of the Planning and Budgeting Committee, by a Grant No. I-1515-303./2019 from the GIF, the German-Israeli Foundation for Scientific Research and Development,  by BSF grant no. 2018204. SSR is grateful to the Aspen Center of Physics for hospitality during initial stages of the project and to the Simons Center for Geometry and Physics.




\bibliography{cat}

\end{document}